\documentclass[useAMS]{mn2e}
\usepackage{times}
\usepackage{rotate}
\usepackage{graphicx}
\usepackage{graphics}
\usepackage{psfig}
\usepackage{lscape}
\input{epsf}
\newif\ifAMStwofonts
\AMStwofontstrue

\def\gsim{\mathrel{\hbox{\rlap{\hbox{\lower4pt\hbox{$\sim$}}}\hbox{$>$}}}}
\def\lsim{\mathrel{\hbox{\rlap{\hbox{\lower4pt\hbox{$\sim$}}}\hbox{$<$}}}}

\title[Fe line cores in AGN]{On the origin of the iron K$\alpha$ line cores in Active Galactic Nuclei}
\author[K. Nandra]{K. Nandra$^{1}$ \\
$^1$Astrophysics Group, Imperial College London, Blackett Laboratory,
Prince Consort Road, London SW7 2AW, UK }
\date{}

\begin{document}

\maketitle
\label{firstpage}

\begin{abstract}
X-ray observations with {\it Chandra} and {\it XMM-Newton} have shown that there are relatively narrow cores to the iron K$\alpha$ emission lines in active galactic nuclei (AGN). Plausible origins for this core emission include the outer regions of an accretion disk, a pc-scale molecular torus, and the optical broad-line region (BLR). Using data from the literature it is shown that no correlation exists between the Fe K$\alpha$ core width and the BLR (specifically H$\beta$) line width.  This shows that in general the iron K$\alpha$ core emission does not arise from the BLR. There is a similar lack of correlation between the width of the Fe K$\alpha$ core and black hole mass. The average K$\alpha$ width is about a factor of two lower than the H$\beta$ width. It therefore seems likely that in many cases the narrow core arises in the  torus. There is a very wide range of observed Fe K$\alpha$ core widths, however, and this argues for multiple origins. The simplest explanation for the observed line profiles in AGN is that they are due to a mixing of very narrow emission from the inner edge of the torus, and broadened emission from the accretion disk, in varying proportions from object-to-object. 
\end{abstract}


\section{INTRODUCTION}
\label{Sec:Introduction}

Iron K$\alpha$ emission lines are extremely common in the X-ray spectra of AGN and provide a potentially unique probe of the circumnuclear environment. The very earliest detections of the emission line were found in heavily obscured objects (e.g. Mushotzky et al. 1978). The emission was attributed to optical broad-line region clouds which, if in the line-of-sight, could simultaneously account for the heavy X-ray absorption (e.g. Holt et al. 1980). 

Later, evidence was found for Fe K$\alpha$ lines in unobscured AGN (e.g. Pounds et al. 1990; Nandra \& Pounds 1994).  These have been attributed to optically thick material out of the line of sight, and are accompanied by a Compton ``reflection" continuum component (Guilbert \& Rees 1988; Lightman \& White 1988; George \& Fabian 1991; Matt et al. 1991). This matter may be identified with the accretion disk (Fabian et al. 1989), or the molecular torus envisaged in orientation-dependent unification schemes for Seyfert galaxies (Krolik \& Kallman 1987; Awaki et al. 1991; Ghisellini, Haardt \& Matt 1994; Krolik, Madau \& Zycki 1994). Iron K$\alpha$ line production in AGN could therefore plausibly originate in three separate sites: the accretion disk, the torus, and the BLR. It should be noted that in the present work we refer to the BLR simply as the region in which the optical broad lines originate, without necessarily identifying it with a physical structure. While the BLR has traditionally been envisaged as a system of ``clouds" , it is possible that it can be identified with a disk wind (e.g. Murray \& Chiang 1995; Elvis 2000), or even the outer regions of the accretion disk (Collin-Souffrin 1987).  

The clearest way to distinguish between the possible origins for the iron K$\alpha$ line is by measurements of its profile, as the various models posit emission from widely differing distances from the central black hole. X-ray spectra from the ASCA satellite (Tanaka, Inoue \& Holt 1994) were the first to have sufficient spectral resolution to address the issue. These spectra supported an origin for the lines in the accretion disk, as heavily broadened and redshifted emission was observed in a number of Seyfert galaxies (Tanaka et al. 1995; Nandra et al. 1997). Data with higher resolution and/or signal-to-noise ratio are now available thanks to the {\it Chandra} and {\it XMM-Newton} satellites. The existence of relativistic accretion disk lines has been confirmed in some cases (e.g. Fabian et al. 2002), and questioned in others (e.g. Gondoin et al. 2001; Reeves et al. 2004), but in general they are difficult to confirm unambiguously, leaving the situation uncertain (Bianchi et al. 2004).  One thing that is clear based on the new data is that relatively narrow ``core'' emission is observed in a large number of objects (Yaqoob et al. 2001; Kaspi et al. 2001). This raises the interesting question of the origin of the core emission, regardless of how often broader lines are seen. 

\begin{table*}
\centering
\caption{Data. 
Col.(1): Object name;
Col.(2): Right Ascension from  NASA/IPAC Extragalactic Database (NED);
Col.(3): Declination (NED);
Col.(4): Redshift (NED);
Col.(5): Black hole mass;
Col. (6) Full width at half maximum of H$\beta$ line km s$^{-1}$;
Col.(7): Fe K$\alpha$ FWHM;
Col.(8): References for mass, H$\beta$ width;K$\alpha$ width. 1) Peterson et al. 2004;
2) Marziani et al. 2003; 3) Yaqoob \& Padmanhaban 2004; 
4) Kaspi et al. 2002 5) Bian \& Zhao 2003; 6) Pineda et al. 1980;
7) Scott et al. 2004; 8) Steenbrugge et al. 2005; 9) Fang et al. 2002; 
10) Zhou \& Wang 2005; 11) Gibson et al. 2005.12) Scott et al. 2005; 
}
\begin{center}
\begin{tabular}{lccccccl}
\hline
Object & RA & DEC & $z$ & $M_{\rm BH}$ & FWHM($H\beta$) & FWHM(Fe K$\alpha$) & References \\
   & (J2000) & (J2000) & & $10^{6} M_{\rm \odot}$ & km s$^{-1}$ & km s$^{-1}$ & \\
  (1) & (2) & (3) & (4) & (5) & (6) & (7) & (8)  \\ \hline
Fairall 9 & 01 23 45.8 & $-58$ 48 20 & 0.047 & $255 \pm 56$ & $6270\pm290$ & $17040^{+55960}_{-14270}$ & 1, 2, 3 \\
3C120 & 04 33 11.1 & $+05$ 21 16 & 0.033 & $55.5 \pm 31.4$  & $2360\pm 170$ & $2000^{+2950}_{-2000}$ & 1, 2, 3  \\
NGC 3516 & 11 06 47.5 & $+72$ 34 07 & 0.009 & $42.7\pm 14.6$ & $3353 \pm 310$ & $1930^{+1380}_{-1380}$ & 1, 1, 3 \\
NGC 3783 & 11 39 0.17 & $-37$ 44 19 & 0.010 & $29.8 \pm 5.4$  & $3570\pm190$ & $1720 \pm 360$  & 1, 2, 4  \\
NGC 4051 & 12 03 09.6 & $+44$ 31 53 & 0.002 & $1.91 \pm 0.78$  & $1072 \pm 112$ & $6330^{+7740}_{-3310}$ & 1, 1, 3 \\ 
NGC 4593 & 12 39 39.4 & $-05$ 20 39 & 0.009 & $5.36 \pm 9.37$ & $5320\pm610$ & $2140^{+8370}_{-1230}$ & 1, 2, 3 \\
MCG-6-30-15 & 13 35 53.8 & $-34$ 17 44 & 0.008 & $1.5 \pm 0.30$ & $1700 \pm 170$ & $3250^{+5230}_{-3250}$ & 5, 6, 2 \\
IC4329A & 13 49 19.2 & $-30$ 18 34 & 0.016 & $9.9 \pm 17.9$ & $5620\pm 200$ & $15090^{+12430}_{-9950}$ & 1, 2, 3 \\
Mrk 279 & 13 53 0.34 & $+69$ 18 30 & 0.030  & $34.9 \pm 9.2$ & $5430\pm 180$ & $4200^{+3350}_{-2950}$ & 1, 2, 7  \\
NGC 5548 & 14 17 59.5 & $+25$ 08 12 & 0.017 & $67.1 \pm 2.6$  & $5830\pm 230$ & $1700\pm 1500$ & 1, 2, 8 \\
H1821+643 &18 21 57.3 & $+64$ 20 36 &  0.297 & ... & $6620\pm 720$ & $11700^{+6400}_{-4100}$ & ..., 2, 9  \\
Mrk 509 & 20 44 09.7 & $-10$ 43 25 & 0.034 & $143 \pm 12$ & $3430\pm 240$ & $2820^{+268-}_{-2800}$ & 1, 2, 3  \\
MR 2251-178 &  22 54 05.8 & $-17$ 34 55 & 0.064 & $102 \pm 20$  &  $6810\pm 460$ & $390^{+260}_{-390}$ & 10, 2, 11 \\ 
NGC 7469 & 23 03 15.6 & $+08$ 52 26 & 0.016 & $12.2 \pm 1.4$ & $2650\pm 220$ & $6310^{+1580}_{-1580}$ & 1, 2, 12 \\
\hline
\hline
\end{tabular}
\end{center}
\end{table*}

The most comprehensive and systematic study of the iron K$\alpha$ line core emission has been presented by  Yaqoob \& Padmanhaban (2004 hereafter YP04) using data from the {\it Chandra} High-Energy Transmission Grating (HETG) spectrometer. The core emission is invariably observed close to the rest energy for neutral iron, i.e. 6.4 keV, implying a very low ionization state for the originating material. The equivalent widths range from a few tens to $\sim 200$ eV, indicating both a large covering fraction, and reasonably high optical depth (say $\tau >0.1$; Awaki et al. 1991; Leahy \& Creighton 1993). The lines are often resolved, but YP04 also presented clear evidence for differences in the velocity width of the cores from source to source, from $1,000-15,000$~km~s$^{-1}$.

A possible explanation for this is that there are varying contributions from the accretion disk and molecular torus (e.g. Reeves et al. 2001; YP04; Zhou \& Wang 2005). If the former dominates, the core is expected to be relatively broad. In the latter case, due to the large distance, it would likely be unresolved even at the HETG resolution.  Another alternative is clearly suggested by the range of line widths, as these fall exactly in the observed range for the optical BLR (e.g., Yaqoob et al. 2001; Bianchi et al. 2003). One might therefore hypothesize that the cores of the iron K$\alpha$ lines originate in that region. The detailed properties of the Fe K$\alpha$ lines might then shed light on the BLR's physical structure. If iron line cores do indeed arise from the BLR there is a clear prediction that the velocity width of the iron lines should be correlated with, and roughly equal to, the width of the optical BLR emission lines. It is the purpose of this {\it Letter} to present a test of this prediction. 

\begin{figure*}
\begin{center}
\rotatebox{270}
{\scalebox{0.65}
{\includegraphics{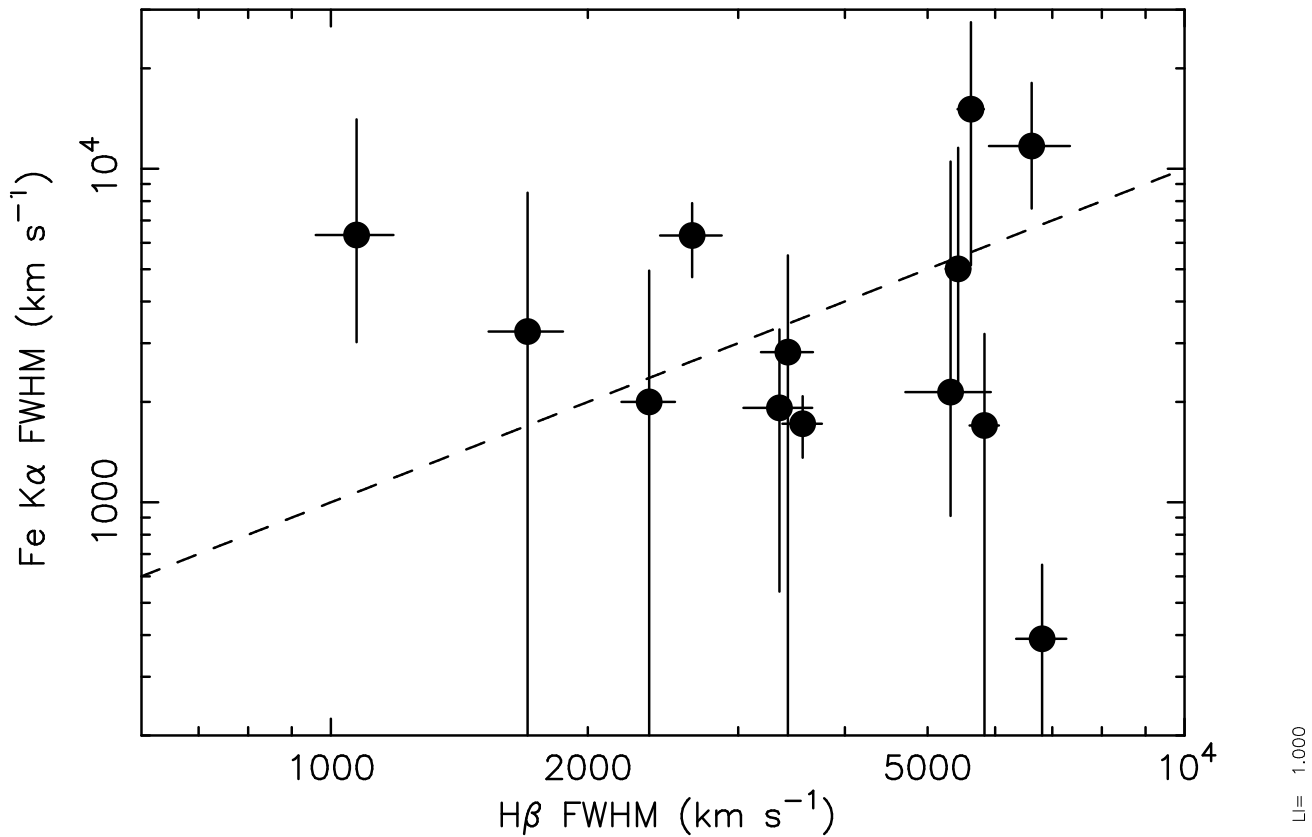}
}}
\caption{Velocity width (FWHM) of the core of the iron K$\alpha$ emission line versus the FWHM of the H$\beta$ line.The dashed line shows a linear 1:1 correlation. The two lines cover approximately the same range  in width, but there is no correlation between them, showing that they have disparate origins.
\label{fig:hbeta}}
\end{center}
\end{figure*}

\section{Data and Results}

To test the prediction of the BLR model for the iron K$\alpha$ line we clearly require profile data in both the X-ray and the optical. The former is more restrictive, as the only instrumental currently capable of resolving lines to sufficient accuracy is the {\it Chandra} HETG. We must also restrict ourselves to objects where it is possible to measure the BLR line width. We have therefore chosen a sample of type I Seyferts for which Fe K$\alpha$ line widths measured with HETG are available in the literature. The resulting sample consists of 14 AGN, listed in Table~1. Our primary resource for the iron line measurements is YP04 and in general we  use those data in preference to others in the literature due to the uniformity of the analysis. Exceptions are NGC 3783 and NGC 5548, where Kaspi et al. (2002) and Steenbrugge et al. (2005) have presented line width measurement based on much longer observations than those presented in YP04. We neglect objects were the line width cannot be constrained by the data due to low signal-to-noise ratio. The errors on the Fe K$\alpha$ width are in general two sided. Following YP04 we adopt the larger of the two-sided errors. In cases where there are multiple observations we take the weighted mean of the measurements and combine the errors.  

All 14 Seyferts have published H$\beta$ width measurements. We adopt Marziani et al. (2003 hereafter M03) as the primary resource for these again for the sake of uniformity, but also because those authors have taken care to deconvolve the broad emission from any narrow components. Uncertainties on the FWHM values were determined from Table~4 of M03, by combining in quadrature the appropriate error on the blue and red wavelengths at half-maximum. Again, the larger of the two-sided errors was then adopted.  In two cases, NGC 3516 and NGC 4051,  we take H$\beta$ widths and uncertainties from Peterson et al. (2004 hereafter P04), noting that these authors measured the line widths from the RMS, rather than the integrated, spectrum. In one case, MCG-6-30-15, no uncertainty is given for FWHM H$\beta$ in the primary reference so we adopt a nominal 10~per cent uncertainty, typical of the upper end of the M03 values. We note, however, that the Fe K$\alpha$ measurement errors are generally much larger, so this assumption, and the adoption of the RMS widths in two cases, should not significantly affect our results. An additional note of caution is that the measurements are non-simultaneous. Any variability  in one width and not the other would introduce scatter into the correlation.

The plot of iron K$\alpha$ versus H$\beta$ width is shown in Fig.~\ref{fig: hbeta}. It is immediately clear from this figure that there is no correlation between the X-ray and optical line widths. Spearman (rank) and Pearson (linear) correlation tests both give statistics with high probabilites ($>50$~per cent) of being observed by chance. While the sample is small, and the measurement errors are relatively large, we can be confident that statistical effects have not contributed to destroying a true correlation between the two quantities. Several things are worth pointing out in this regard. Firstly, the object with the lowest FWHM H$\beta$,  NGC 4051 (which is a so-called Narrow-Line Seyfert 1; NLS1 Osterbrock \& Pogge 1985) has a relatively broad iron K$\alpha$ line core. The broadest H$\beta$ line object is seen in MR 2251-178, but this has the narrowest Fe K$\alpha$ core in the sample, and its value is relatively well determined.  Furthermore, removing either (or indeed both) of these extreme objects from the sample fails to reveal a correlation. Finally, it is clear that objects with similarly broad optical lines have  Fe core widths which are very different. For example, the widths in MR 2251-178 and H1821+643, the two objects with the broadest optical lines (FWHM H$\beta > 6500$ km s$^{-1}$), have K$\alpha$ widths spanning from the very narrowest to the very broadest observed values. 

\begin{figure}
\begin{center}
\rotatebox{270}
{\scalebox{0.3}
{\includegraphics{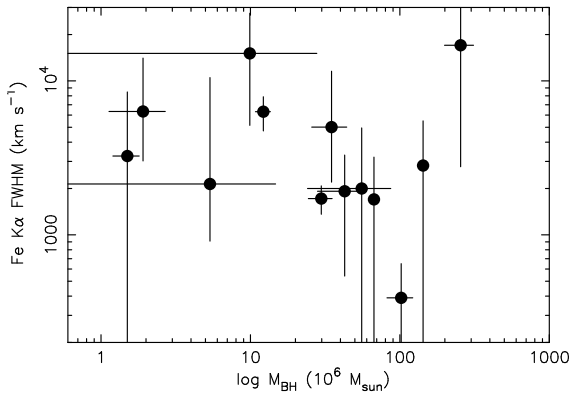}
}}
\caption{Velocity width (FWHM) of the core of the iron K$\alpha$ emission line versus the black hole mass. There is no apparent relationship between these quantities. This and Fig.~\ref{fig:hbeta} also demonstrate the wide variation in Fe K$\alpha$ FWHM from object-to-object (see also YP04). 
\label{fig:mass}}
\end{center}
\end{figure}

Fig~\ref{fig:mass} shows the relationship between the iron K$\alpha$ line width and the black hole mass. The latter were taken largely from the compilation of O'Neill et al. (2005), which mostly originate in P04. Primary references are given in Table 1. For masses not taken from P04 we assign a nominal 20~per cent error. It is clear both that the FWHM measurements are not consistent with a constant, are not correlated, and nor are they consistent with $M \propto v^{2}$. The first is expected if the line arises from the same radius with respect to the gravitational radius (i.e. $r/r_{\rm g}$ is constant) and the latter two if the line arises from the same physical radius (i.e. $r$ is constant). 

\section{DISCUSSION}

The Fe K$\alpha$ core and BLR line widths in AGN differ substantially from object-to-object and cover a wide range. We have shown in the above analysis, however, that there is no relationship between the two. The most immediate conclusion is therefore that the Fe K$\alpha$ core does not generally arise from the optical BLR. 

In the above analysis, we have concentrated exclusively on the H$\beta$ width as a measure of the BLR velocity. It has become well established from reverberation mapping experiments (e.g. Clavel et al. 1991; Peterson et al. 1991) that there is not a single ``BLR". Indeed this work has shown that the broad optical lines in AGN arise from a range of radii in rough proportion to the ionization potential of the species. One can image the BLR either as a series of clouds, or a continuous structure such as a wind, with a range of ionization states as a function of radius. Because iron is likely to survive at some level in all of the material responsible for the various broad lines, Fe K$\alpha$ emission in the BLR should occur at all radii, with the overall width representing an optical-depth weighted average over the whole region. As H$\beta$ is emitted at relatively large radii, one would expect the Fe K$\alpha$ cores to be, on average, broader than H$\beta$. There is no evidence for this, and if anything the latter is the case. The weighted mean H$\beta$ width for the sample is $3200 \pm 60$~km$s^{-1}$, whereas the Fe K$\alpha$ line is less than half that width, with weighted mean FWHM $1350 \pm 250$. The latter value is in agreement with that presented by YP04.  

The implication is that, typically, the Fe K$\alpha$ core comes from {\it outside} the BLR and hence is plausibly identified with the torus responsible for obscuring the BLR in the unification schemes (Antonucci \& Miller 1985).  A torus origin for the neutral Fe K$\alpha$ line is thought to be most likely in Seyfert 2 galaxies. The line cores are unresolved at the HETG resolution  (e.g. Sambruna et al. 2001; Ogle et al. 2003) but there is a ``Compton shoulder'' which shows unambiguously that they originate in optically thick material (Iwasawa et al. 1997; Bianchi et al. 2002; Matt et al. 2004). A tentative detection of the Compton shoulder has been made in the Seyfert 1 NGC 3783 (Kaspi et al. 2002; Yaqoob et al. 2005), supporting a torus origin for the line core in that object also. 

If the line originates universally from the inner edge of the torus one {\it still} might expect the line widths to be correlated, as the H$\beta$ emission radius and the inner edge of the torus (which may represent the dust sublimation radius) should both be controlled by the power of the central source. Alternatively, the inner edge of the torus might occur at a fixed number of gravitational radii. In the latter case, however, one would predict a constant Fe K$\alpha$ core width. The details of this are, however, strongly dependent on the precise geometry and physical parameters, because the iron K$\alpha$ emission is expected to arise from a range of radii in the torus.  It seems clear, however, that in objects where the iron line core width exceeds the H$\beta$ width, one can rule out the torus as a possible origin. While the inner BLR remains a possible site for line emission, in these objects the Fe K$\alpha$ emission core seems most likely to be associated with the outer regions of the accretion disk, given the compelling evidence for the origin of the broadest Fe K$\alpha$ emission in the disk (e.g. Vaughan \& Fabian 2004; Ponti et al. 2004). 

Three notes of caution should be sounded about the analysis described above. Firstly, the sample size is currently very small. It will clearly be important to see if the conclusions remain robust when more objects are included. This requires further high signal-to-noise ratio spectra of Seyferts to be obtained with {\it Chandra}. Secondly, it is well known in spectroscopy that the continuum parameterization can affect the line parameters significantly. These effects are exacerbated when the signal-to-noise ratio is low, as is the case with some of the spectra of Seyferts used in this {\it Letter}. When considering the narrow core of the line and where high resolution data are available, however, the uncertainties due to continuum modeling should be minimised. Finally, as noted earlier, the Fe K$\alpha$ and H$\beta$ measurements are not simultaneous, hence variability of the widths could introduce scatter into their relationship. A particularly well-studied case is that of NGC 5548. P04 have presented a series of measurements of the H$\beta$ width, which show excess RMS variability of $\sim 15$~per cent. This is larger than the typical measurement error in FWHM H$\beta$, but the uncertainties in the X-ray line widths still dominate.
Variability should not therefore strongly affect our conclusions, but obtaining strictly simultaneous, or at least contemporaneous, X-ray and optical spectroscopy would clearly be highly desirable. 

Overall it appears that, based on the present data, the structures responsible for the production of the optical broad lines (e.g. clouds, disk wind) are a negligible source of iron K$\alpha$  emission in Seyfert galaxies. This may not be at all surprising given the physical parameters expected for the BLR. Assuming ``typical'' values for BLR clouds of $N_{\rm H} \sim 10^{23}$~cm$^{-2}$ and a covering fraction of 10~per cent one would predict an equivalent width of order only 10 eV for the Fe K$\alpha$ line (Leighy \& Creighton 1993). Similarly, the column densities in the outflows responsible for the UV and X-ray absorption lines in Seyferts (e.g. Mathur, Wilkes \& Elvis 1998; Crenshaw, Kraemer \& George 2004) are typically $\sim 10^{22}$~cm$^{-2}$.  The optical depth of this outflowing gas,  which may be identified with the disk wind, are therefore too low to produce significant iron K$\alpha$ emission even if the covering fraction is high. In individual cases it will clearly  be very difficult to rule out a BLR contribution to Fe K$\alpha$. Our conclusion, however, is that the simplest approach towards the modeling the iron K$\alpha$ line complex in Seyferts is to assume a mix of contributions from both the torus and accretion disk, in differing proportions depending on the source. The key questions that now remain relate to the reasons why those proportions differ and, particularly, why in some cases one or other of the components of the iron K$\alpha$ line appear to be absent. 

\section*{Acknowledgements}

The author gratefully acknowledges the support of the Leverhulme Trust in the form of a research fellowship. I am indebted to Ian George and Brad Peterson for helpful discussions. I thank the referee, Kazushi Iwasawa, for constructive comments. This research has made use of the NASA/IPAC Extragalactic Database (NED) which is operated by the Jet Propulsion Laboratory, California Institute of Technology, under contract with the National Aeronautics and Space Administration and of the High Energy Astrophysics Science Archive Research Center (HEASARC), provided by NASA's Goddard Space Flight Center.


\bsp

\label{lastpage}

\end{document}